\pdfoutput=1 
\documentclass[a4paper,USenglish,cleveref,autoref,thm-restate]{lipics-v2021}

\hideLIPIcs  

\title{Automatic Test-Case Reduction in Proof Assistants: A Case Study in Coq}
\titlerunning{Automatic Test-Case Reduction in Coq}

\author{Jason Gross}{CSAIL, Massachusetts Institute of Technology, 77 Massachusetts Ave., Cambridge, MA 02139, USA \and MIRI, USA \and \url{https://jasongross.github.io/} }{jgross@mit.edu}{https://orcid.org/0000-0002-9427-4891}{}

\author{Théo Zimmermann}{Inria, Université de Paris, CNRS, IRIF, F-75013, Paris, France \and \url{https://www.theozimmermann.net} }{theo@irif.fr}{https://orcid.org/0000-0002-3580-8806}{}

\author{Rajashree Agrawal}{Reed College, 3203 SE Woodstock Blvd, Portland, OR 97202, USA
}{ragrawal@reed.edu}{https://orcid.org/0000-0001-7617-9180}{}

\author{Adam Chlipala}{CSAIL, Massachusetts Institute of Technology, 77 Massachusetts Ave., Cambridge, MA 02139, USA \and \url{http://adam.chlipala.net/} }{adamc@csail.mit.edu}{https://orcid.org/0000-0001-7085-9417}{}

\authorrunning{J. Gross and T. Zimmermann and R. Agrawal and A. Chlipala} 

\Copyright{Jason Gross and Théo Zimmermann and Rajashree Agrawal and Adam Chlipala} 



\category{} 

\relatedversion{} 
\relatedversiondetails[
  cite=coqpl-15-coq-bug-minimizer]{Earlier}{https://jasongross.github.io/papers/2015-coq-bug-minimizer.pdf} 

\supplement{\url{https://figshare.com/s/60037cb0c9bbf464e686}}

\nolinenumbers 

\EventEditors{John Q. Open and Joan R. Access}
\EventNoEds{2}
\EventLongTitle{42nd Conference on Very Important Topics (CVIT 2016)}
\EventShortTitle{CVIT 2016}
\EventAcronym{CVIT}
\EventYear{2016}
\EventDate{December 24--27, 2016}
\EventLocation{Little Whinging, United Kingdom}
\EventLogo{}
\SeriesVolume{42}
\ArticleNo{23}

\usepackage{usebib}
\newbibfield{title}
\bibinput{coq-bug-minimizer}

\makeatletter
\newcommand\autorefs[1]{\@first@ref#1,@}
\def\@throw@dot#1.#2@{#1}
\def\@set@refname#1{
    \edef\@tmp{\getrefbykeydefault{#1}{anchor}{}}%
    \xdef\@tmp{\expandafter\@throw@dot\@tmp.@}%
    \ltx@IfUndefined{\@tmp autorefnameplural}%
         {\def\@refname{\@nameuse{\@tmp autorefname}s}}%
         {\def\@refname{\@nameuse{\@tmp autorefnameplural}}}%
}
\def\@first@ref#1,#2{%
  \ifx#2@\autoref{#1}\let\@nextref\@gobble
  \else%
    \@set@refname{#1}
    \@refname~\ref{#1}
    \let\@nextref\@next@ref
  \fi%
  \@nextref#2%
}
\def\@next@ref#1,#2{%
   \ifx#2@, and~\ref{#1}\let\@nextref\@gobble
   \else, \ref{#1}
   \fi%
   \@nextref#2%
}
\makeatother

\makeatletter
\newcommand{\todo}[1]{%
\@latex@warning{TODO: \detokenize{#1} on page \thepage}%
\textcolor{red}{[\textbf{TODO:} #1]}}%
\makeatother
\newcommand{\coqbug}[1]{\href{https://github.com/coq/coq/issues/#1}{coq/coq\##1}}

\begin{document}

\maketitle

\begin{abstract}
  As the adoption of proof assistants increases, there is a need for efficiency in identifying, documenting, and fixing compatibility issues that arise from proof assistant evolution.
  We present the Coq Bug Minimizer, a tool for \emph{reproducing buggy behavior} with \emph{minimal} and \emph{standalone} files, integrated with coqbot to trigger \emph{automatically} on Coq reverse CI failures.
  Our tool eliminates the overhead of having to download, set up, compile, and then explore and understand large developments: enabling Coq developers to easily obtain modular test-case files for fast experimentation.
  In this paper, we describe insights about how test-case reduction is different in Coq than in traditional compilers.
We expect that our insights will generalize to other proof assistants.
  We evaluate the Coq Bug Minimizer on over 150 CI failures.
  Our tool succeeds in reducing failures to smaller test cases in roughly 75\% of the time.
  The minimizer produces a fully standalone test case 89\% of the time, and it is on average about one-third the size of the original test.
  The average reduced test case compiles in 1.25 seconds, with 75\% taking under half a second.
\end{abstract}

\section{Introduction}

In the world of machine verification, the dream is to prove the correctness of every program.
Projects such as Coq Coq Correct!~\cite{coq-coq-correct} make significant progress towards this dream for even our most foundational tools: proof assistants themselves.
However, large swathes of proof-assistant software—such as tactic languages, elaboration hints, and document managers—remain unproven, lacking even adequate test-suite coverage!

As a solution to expanding the test-suite coverage for the proof assistant Coq, developers adopted ``reverse'' continuous integration (CI)~\cite{zimmermann:tel-02451322,ochoa2022breakbot} wherein changes in Coq are tested against a crowdsourced suite of external Coq projects maintained by different teams in different repositories.
In this manner, user-centric concerns are well addressed.
To prevent the crowdsourced test suite from shrinking, when Coq evolves in a desired direction but breaks some external project in the process, developers of Coq will fix the compatibility issue in the external project.
\emph{We believe that to facilitate the use of proof assistants in industry-scale projects, it is essential to make it easy to find, understand, and fix compatibility issues as the proof assistant continues to evolve.}

Since the external projects in the Coq test suite are large and intricate, debugging and fixing failures reported by the reverse CI is a time- and effort-intensive process for developers.
They must perform many steps before beginning to understand and work on the bug.
First, developers will tediously sit through the process of downloading, setting up, and compiling the external project.
Then, they may have to take on the daunting task of figuring out the larger project context which is not even directly relevant to the bug!

The current debugging process can be significantly optimized for developer experience.
Additionally, the current process does not easily yield test cases to add to Coq's internal test suite.
Instead the test cases remain buried in external developments whereas we would like to bring bugs to the center!
In order to improve the debugging process, we built the Coq Bug Minimizer%
\footnote{%
Available on GitHub in \href{https://github.com/JasonGross/coq-tools}{\texttt{JasonGross/coq-tools}}%
}
which \textbf{reproduces buggy behavior} in \textbf{minimal} and \textbf{standalone} files.
Typically, minimized files reduce the total number of lines of code involved in exhibiting buggy behavior by about a factor of three, making it significantly easier for developers to observe, play with, understand, and fix bugs.
Furthermore, we have integrated the Coq Bug Minimizer with coqbot~\cite{zimmermann:hal-03479327} to trigger \textbf{automatically} on reverse-CI failures, reducing the friction of building minimized files.

Test-case reduction has already a rich literature~\cite{chen_survey_compiler_testing}. However, it is focused mostly on traditional languages such as C, and even generic reduction techniques may not apply so well to proof assistants.
In this paper, we share what we have learnt about where test-case reduction is harder and where it is easier in Coq than in traditional compilers, and describe how we got around the difficulties.
Drawing on empirical results from nearly a year of use in Coq's production CI system, we reflect on how effective our style of test-case reduction has been and where the biggest opportunities for improvement remain.
We believe that our methods may be of interest for developers of other proof assistants who are also facing a tradeoff between enabling evolution and preserving stability, in a context of industrial use.

For the mobile reader, \autoref{sec:example} introduces a constructed example of test-case reduction in Coq, and articulates desiderata for test-case reduction in the proof-assistant setting.
\autoref{sec:easier} details aspects of traditional-setting test-case reduction that are simpler or irrelevant in Coq.
Then \autorefs{sec:error-messages,sec:minimal,sec:standalone,sec:smooth-dev-experience} explore the four desiderata and describe the details of our solution to the more important challenges of each.
\autoref{sec:alt-usage} forays into the applicability of the Coq Bug Minimizer for bug reporter workflow as a secondary use case.
Finally, \autoref{sec:evaluation} presents our deployment in Coq's production CI, with analysis of how effectively different test cases were minimized; \autoref{sec:related-work} describes connections to related work; and \autoref{sec:future-work} discusses our thoughts on the most worthwhile improvements to make to our tooling.

\section{Desiderata}\label{sec:example}

To add color to our picture, let's begin with a constructed example of minimizing a reverse-CI failure.
Our objective is to explore the space of file modifications that will aid human understanding of the bug.
Consider the following Coq source file:
\begin{verbatim}
Require Import UsefulTactics.
Definition zero := 0. Definition one := 1.
Definition two := 2. Definition three := 3.
Lemma foo : forall x, x = zero -> S x = one.
Proof. crush. Qed.
\end{verbatim}

Suppose the \texttt{crush} tactic triggered a new bug in Coq.
The most obvious move is to find deletable sentences and delete them, producing a smaller file:
\begin{verbatim}
Require Import UsefulTactics.
Definition zero := 0. Definition one := 1.
Lemma foo : forall x, x = zero -> S x = one.
Proof. crush.
\end{verbatim}

The file still depends on an imported module not native to the Coq standard library.
The next move is to inline this dependency, producing a standalone file:
\begin{verbatim}
Module UsefulTactics.
Ltac head expr := match expr with | ?f _ => head f | _ => expr end.
Ltac head_hnf expr := let expr' := eval hnf in expr in head expr'.
Ltac crush := intros; subst; try reflexivity.
End UsefulTactics.
Import UsefulTactics.
Definition zero := 0. Definition one := 1.
Lemma foo : forall x, x = zero -> S x = one.
Proof. crush.
\end{verbatim}

Now we may look for any more opportunities to delete lines, producing a standalone, reduced file:
\begin{verbatim}
Ltac crush := intros; subst; try reflexivity.
Definition zero := 0. Definition one := 1.
Lemma foo : forall x, x = zero -> S x = one.
Proof. crush.
\end{verbatim}

From the above process we can extrapolate desiderata for the Coq Bug Minimizer.

\begin{enumerate}
\item \textbf{Reproducing buggy behavior:} Deciding when two source files indicate the same bug.
Many reasonable file simplifications lead to incidental changes in error messages.
The Coq Bug Minimizer must tradeoff between preserving specific details of error messages and aiding human understanding of the underlying bug.

\item \textbf{Minimal files:} Exploring the space of program simplifications in a smart way with respect to constraints of the proof-assistant setting.
Many research papers in the software-engineering community have been written on just this topic~\cite{Cleve2000,zeller2009programs,Burger2005,delta,Zeller2002}, but constraints in a proof-assistant setting are uncommon in conventional programming.
For instance, highly automated Coq developments often have long compile times even for single files, so we may need to be more frugal in how many program variants we test.

\item \textbf{Standalone files:} Creating standalone files that illuminate new test cases and can be added to Coq's internal test suite.
This is difficult in dependently typed languages with metaprogramming facilities such as Coq.
For instance, eliminating needless dependencies in simply typed languages may be trivial, but dependently typed languages eliminate the distinction between runtime and compile time resulting in tight coupling between files.

\item \textbf{Smooth developer experience:} Automatically finding which file triggered a bug, with which compilation settings, including path information to find dependencies.
The Coq Bug Minimizer must work with the wide variety of build systems used in different Coq libraries.

\end{enumerate}

Achieving each desideratum posed interesting challenges, and required making several design choices.
Before proceeding to share solutions to these challenges, we note the ways in which test-case reduction is \emph{simpler} in the proof-assistant setting than in other settings.

\section{Simplifications of the Proof-Assistant Setting}\label{sec:easier}

Classic delta debugging~\cite{Zeller2002} is a technique in test-case reduction for traditional compilers.
It employs binary search through program structure to find subprograms that can be removed while preserving properties relevant to triggering specific bugs for the chosen compiler.
Coq's lack of forward references permits a simpler method: first remove everything after the error-message-generating line, and then try removing the syntactic units beforehand in-order, one-at-a-time.
Unlike in languages from Java to Haskell, where all functions in a file are considered mutually recursive, in Coq there should be no way for one error-message-generating line of a file to change behavior based on modifications to later lines.
In this manner we reduce the number of ``experiments'' on program variants, which is especially useful when each program variant requires significant processing time as is often the case in Coq.

Our empirical evaluation (\autoref{sec:evaluation}) demonstrates that this strategy is adequately performant.
We conjecture that the reason for this adequate performance is that dependency trees of Coq theorems and proofs tend to be relatively deep compared to the number of definitions and theorems in any single file.
This hypothesis is borne out by the fact that our typical ``minimal'' test case tends to be only about a third the size of the total amount of code in all files in the dependency tree of the initial test case.
If instead there were orders of magnitude more useless lines than true dependent lines, we expect that a binary-search strategy would be required for adequate performance.

\section{Reproducing Buggy Behavior}\label{sec:error-messages}

How do we know modifications to source files are genuine simplifications that have not masked bugs?
What does it mean to reproduce the ``same'' bug?
We generate a file that succeeds on the previous version of Coq and continues to fail on the modified version of Coq, with the same error message that showed up in the reverse CI.
However, the error message of the generated file does not need to be \emph{exactly} the same as in the original file, so long as the reason for the error message is the same.
Thus, we modify our goal to reproducing buggy behavior in place of reproducing the ``same'' bug.

We apply the following relaxations in comparing error messages.
\begin{enumerate}

\item Universe inconsistencies are how Coq prevents users from proving absurdity by assuming a ``set of all sets.''
The explanations of universe inconsistencies in error messages are sensitive to how many universes are floating around and in what order constraints were added.
Rather than requiring output files to mimic the error messages exactly, we only require that they result in \emph{some} universe inconsistency.

\item Any two error messages about ``forgotten universes'' are considered matching, since these tend to arise only from very specific Coq internal errors.

\item Usually differences in numbering, e.g.\ in universes or autogenerated identifiers, are incidental and are not treated as implying different error messages.  One special case is lengths of universe instances, so we look for the text ``Universe instance should have length'' in the error message and only use number-insensitive comparison if this text is not found.

\item We consider any error messages containing ``Unsatisfied constraints: \ldots\ (maybe a bugged tactic)'' as equivalent, since related bugs are localized to one relatively small part of the Coq implementation, and small changes to a source file can modify constraints significantly.

\item We also ignore filenames, line references, and word wrapping in comparing error messages.

\end{enumerate}

\section{Minimal Files}\label{sec:minimal}

Test-case reduction is powerful in making long source files more comprehensible to developers.
In addition to this, external projects in Coq can take minutes or hours to compile, so the edit-compile-test-debug loop is long.
We have two additional goals to improve this workflow.
\begin{enumerate}
\item Finding minimal test cases as fast as possible, given that experimenting with each program variant has long compilation time.
\item Compilation of the test case in seconds or fractions of a second so that developers can fluidly try hypotheses for solutions.
\end{enumerate}

\subsection{Making the Minimization Process Itself Fast}

In our goal to get the shortest reproducing test case as quickly as possible, it helps to first make any changes that might significantly speed up the execution time, and only after we're done with all of the changes that might improve running time should we try to further minimize the file with changes that are unlikely to impact compile time.

The slowest part of almost all Coq developments is proof scripts.
Hence we attempt to remove proof scripts as early as possible.
Since proof assistants check that proofs are valid, we cannot simply remove a proof, like we might remove a function body in a traditional programming language.
However, most proof assistants have some mechanism for ``giving up'' on a proof or ``trusting'' the user, and Coq is no exception.
Its mechanism involves any of \verb|Admitted|, \verb|Admit Obligations|, or the \verb|admit| tactic.
Replacing proof blocks with these commands, rather than just removing proof scripts, allows us to make much smaller and faster examples than might otherwise be possible.

\subsection{Finding Textually Smaller Test Cases}

The simplest function of the bug minimizer is to remove unneeded lines.
As noted in the prior section, we try removing one syntactic unit at a time, moving backwards from the unit that triggered the error message.

However, we can easily enough get stuck in local minima, when we remove single commands and check that bug behavior is unchanged.
For instance, there may be an irrelevant lemma that we want to remove.
\begin{verbatim}
Lemma irrelevant : two = 2.
Proof. reflexivity. Qed.
\end{verbatim}
Since Coq forbids nested lemmas, removing statements one-at-a-time will not work, as the state
\begin{verbatim}
Lemma irrelevant : two = 2.
Proof. reflexivity.
\end{verbatim}
results in an error about nested proofs, if there is a theorem afterward.

We instead group statements into \emph{definition} blocks to be removed all at once.
We get information about definitions by parsing the output of \texttt{coqtop -emacs -time}.
This way, we can remove the lemma block all at once.

We could in theory deal with more complicated nesting structure, for example trying to remove an entire section or module at a time.
The delta tool~\cite{delta} is in fact built around preprocessing the file into one that exposes nested structure clearly, then removing well-parenthesized blocks.
However, removing statements, grouped into definitions as necessary, sufficies for removing time-consuming code.

\subsubsection{The Program Construct}
One Coq construct that does not fit neatly into this approach is \verb|Program|, where a function definition is associated with following proofs of obligations related to dependent typing.
We cannot just look for \verb|Program| statements followed by \verb|Obligation| blocks to remove all together, because \verb|Obligation| blocks can be interleaved with other definitions.
Luckily, we can replace any obligation block with a use of the \verb|Admit Obligations| command, which admits all remaining obligations---and it happily handles any case with \emph{no} remaining obligations, so we need not worry about introducing duplicate invocations.

\subsubsection{Empty Sections and Modules}
Removing statements one-at-a-time will not always be able to remove empty sections (nor empty \verb|Module|s or \verb|Module Type|s).
That is why we have a pass dedicated to removing empty sections, modules, and module types.

\subsubsection{Exporting Modules}
Coq's features to import and export modules (e.g., including all definitions of one module inside another) can create some particularly thorny situations for statement-at-a-time shrinking.
If we remove just an \verb|Import| commands, then later commands fail because important identifiers are out-of-scope.
If we remove just the definition of the imported module, then the \verb|Import| fails.
The solution is to merge these two commands, so that they become a candidate for removal together.
We change \verb|Module| commands into \verb|Module Export| commands to this end.
Often that change renders later \verb|Import| commands redundant, so they are removed by later passes.

\subsubsection{Splitting Definitions}
One pass in the minimizer tries to replace traditional definitions with uses of the interactive proof mode, which is a first step toward admitting those proof bodies (i.e., postulating existence of identifiers rather than giving their definitions) in later steps.

\subsubsection{Early Removal of Unused Constants}
There are some likely-to-succeed steps that we try early on, which are superseded by removing each and every structured block one at a time but may result in faster minimization.
The primary example of this sort of step is removing tactics, \verb|Variable| and \verb|Context| statements, and definitions which are not referred to at all after their definitions.

\subsubsection{Splitting \texttt{Import} and \texttt{Export}s}\label{sec:split-imports}
It may be the case that users import modules that they never use, such as in \texttt{Import unused1 used unused2}.
To allow eventual removal of \verb|unused1| and \verb|unused2| even when the \verb|Import used| statement cannot be removed, we have a pass that attempts to split such statements into separate \verb|Import| statements, resulting in \texttt{Import unused1. Import used. Import unused2.}

\subsection{Finding Test Cases That Coq Processes More Quickly}

We mentioned how admitting proofs is a very handy step to shrink files and get them processed more quickly.
There are, however, a few gotchas to keep in mind.

The first quirk is around transparency vs. opacity of lemma definitions; that is, whether the generated proof term is accessible to later definitions.
Either choice (transparent vs. opaque) can break some developments.
Marking a proof-mode definition \emph{opaque} will break later definitions that unfold the definition and then perform further tactic-based surgery on it, while marking a proof-mode definition \emph{transparent} could cause previously failing unfoldings to succeed.
Therefore, we always try both styles of marking a lemma admitted.

Some lemma proofs are declared as transparent rather than opaque, where later steps really do depend on their details.
If those dependencies are too specific, then our shrinking heuristics are not going to work well.
However, one common-enough case is where a later definition uses tactics to \emph{unfold} an earlier definition, going on to use other tactics that may very well be able to adapt to changes in that definition.
There are at least two different ways to mark a proof as admitted (\texttt{Admitted} vs.\ using a preexisting \texttt{Axiom}), which can switch up whether the associated definition is considered transparent or opaque.

Additionally, we may want to admit some parts of the proof script without replacing all of it.
Currently, we use a rather conservative heuristic:
Coq has a tactical \verb|abstract| that executes the tactic it is passed as an argument, making the resulting proof term opaque.
Such subproofs should be able to be replaced with \verb|admit| without changing the behavior of the proof script.
The details are a little subtle, e.g.\ to avoid changing which section variables a proof depends on and thus changing its type outside the section.

\section{Standalone Files}\label{sec:standalone}

While the complex structure of external developments is a boon to stress-testing Coq, there are three reasons for wanting to reproduce bugs in standalone files.
\begin{enumerate}
\item It is challenging for developers to understand the intricacies of external developments well enough to diagnose root causes.
\item  Build systems are necessary to handle multiple files, but using them adds unnecessary overhead in the debugging workflow.
\item Intricate file-dependency structure complicates test-suite infrastructure, whereas having self-contained files results in a simpler test suite.
\end{enumerate}

Na\"ively, the way to produce a standalone file is to \emph{linearize} the dependency tree and combine the contents of all files.
We saw an example of roughly this strategy in \autoref{sec:example}, and e.g.\ C compilers follow this strategy in preprocessing \verb|#include| statements.

Two difficulties arise when following this strategy in Coq:
\begin{enumerate}
\item As in all languages that allow shadowing of global symbols, inlining files changes what names are available and hence may result in unintended changes of behavior.
The dependent typing and metaprogramming facilities of Coq largely eliminate the distinction between runtime and compile time.
As a result, we have to inline not just function declarations but also function bodies, and thus the problem of name resolution is comparatively harder in Coq and similar languages than in those with simple types and without metaprogramming facilities.
Furthermore, Coq has additional quirks around name resolution and (lack of) namespacing that have to be managed and worked around.
\item Coq has a great deal of global state (e.g., notations, universe polymorphism, the default tactic mode) that changes the way sentences are interpreted.
Because there is no way to isolate changes on this global state fully, there may not even be \emph{any} linearization that reproduces the same behavior.
\end{enumerate}

\subsection{Addressing Shadowing and Name Resolution}\label{sec:name-resolution}
Coq assigns names based on three components: the name and location of the file in which the identifier is defined, the module structure surrounding the identifier, and the final name.
For example, the constant \texttt{Coq.MSets.MSetPositive.PositiveSet.t} is defined in the file \texttt{MSets/MSetPositive.v}, which is bound to \texttt{Coq.MSets.MSetPositive}, in the module \texttt{PositiveSet}, with the name \texttt{t}.

If we were to inline this file into some other file \texttt{bug.v}, then the constant becomes \texttt{bug.PositiveSet.t}.
We now have two choices: we can attempt to adjust the name of the constant on inlining, or we can adjust references to the constant.

We combine these strategies to maximize the chance of succesfully inlining dependencies.

First, as shown in the example in \autoref{sec:example}, we wrap the contents in a module whose name matches that of the file (in this case, we wrap the contents in \texttt{Module MSetPositive}).
Furthermore, since users can refer to this constant as \texttt{Coq.MSets.MSetPositive.PositiveSet.t}, \texttt{MSets.MSetPositive.PositiveSet.t}, or \texttt{MSetPositive.PositiveSet.t}, we can wrap this module in further modules (\texttt{Coq} and \texttt{MSets}) and \texttt{Export} them to make this naming scheme available.
Finally, because Coq forbids multiple modules with the same absolute kernel name, we must wrap the top-level module in yet another module, with a uniquly generated identifier.
While this strategy is not perfect, running afoul of \coqbug{14587} for example, we try a couple of variations on this strategy, and very often one of them is adequate for reproducing buggy behavior.

Second, we want to adjust references so that they still point at the same underlying object after inlining.
Coq helpfully emits \emph{globalization} files, which contain information about how Coq resolves almost all names in the file.
Since Coq generates and installs these \texttt{.glob} files, we can use this information to transform both the names in the files we inline and the names that refer to constants in that file.

However, we cannot just blindly update all names, because these \texttt{.glob} files are not perfectly accurate\footnote{See \coqbug{15497} and \coqbug{14537}.} and are not complete\footnote{They are missing information, for example, on tactic-name resolution and notation interpretation.}.
Instead, we have found in practice that the most important names to resolve are those used in \texttt{Require}, \texttt{Import}, and \texttt{Export} statements.
\texttt{Require} statements are sensitive to the searchpath flags (\verb|-Q| and \verb|-R|) passed to Coq.
If we are inlining a file from Flocq into a file from VST, for example, the \texttt{Require}s in the Flocq file may not resolve to the same files on disk when compiling with the compiler flags that VST uses.
\texttt{Import} and \texttt{Export} statements, while not dependent on searchpath flags to the same extent as \texttt{Require}, still seem empirically more likely to refer to potentially ambiguous names than most other statements.
Hence we choose to resolve the names used in \texttt{Require}, \texttt{Import}, and \texttt{Export} statements when inlining, letting Coq determine all other name resolution.

\subsection{Addressing Nonlinearizability of Global State}\label{sec:linearize-global-state}
While shadowing and name resolution are mechanically resolvable at least in theory, the global state of Coq is sufficiently disorganized that we are not aware of any fully general technical means of linearizing Coq files.\footnote{%
The \texttt{Require} command results in many side effects, including global setting of flags, opacity, and argument status; behavior of \texttt{auto with *}; hint databases; global overwriting of Ltac definitions; presence or absence of constants that change the behavior of built-in tactics such as \texttt{tauto}; and even the presence of constants with certain kernel names can change shadowing behavior.
Some of these effects can even be set on the command-line, and at present there is no way to determine what flags were used to compile a given installed file.
}
Hence our approach here consists of several partial workarounds.

The most basic technique to attempt to isolate global state is to wrap the inlined file in a module.
Most state not explicitly marked as \texttt{Global} does not escape the boundaries of the module it is defined inside.
As we already use module wrapping to handle name resolution as discussed in \autoref{sec:name-resolution}, we already reap the benefits of this technique.

Our only other technique is to try multiple linearizations and hope that one of them is adequate.
We try inserting the file being inlined at the top of the file, as well as at the location where it is \texttt{Require}d.
In the future, we might also want to try moving \texttt{Require}s up higher in the file, to try to handle more situations.

In \autoref{sec:future-work}, we discuss a few potential future avenues to better handling of global state.
For example, we may want to more explicitly manage the state before and after inlining a file by taking advantange of Coq's ability to print the current settings of flags with \texttt{Print Options}.

\subsection{Getting to Standalone Files Quickly}
We have a flag that allows inlining dependencies all at once, much like \texttt{gcc} inlines all \verb|#include|d files at once.
While originally all files were minimized in that way, having to process such a large file slowed down minimization drastically, often resulting in minimization times of multiple weeks.
As a result, the current default behavior is to minimize the current file before inlining other files.

Futhermore, we want to ensure that we only inline files that are actually used.
Much like we want to split \texttt{Import} and \verb|Export| statements in \autoref{sec:split-imports}, we also want to split \verb|Require| statements, for example from \texttt{Require unused1 used unused2.} to \texttt{Require unused1. Require used. Require unused2.}

Additionally, if the buggy behavior depends on a file only for its own dependencies, we prefer to inline the transitive dependency directly rather than needing to inline the entire intermediate file.
To that end, we have a pass that performs the transitive closure of the dependency relation, inserting \verb|Require| statements at the top of the file for all transitive dependencies of the file being minimized.
Because we insert the \verb|Require|s in dependency order, removing one statement at a time in reverse order will give us the minimal \verb|Require|s needed to reproduce the error message.
This strategy ensures that we only inline dependencies that are actually necessary.

\section{Smooth Developer Experience}\label{sec:smooth-dev-experience}

In order to analyze a specific source file, we need to take a few steps.
\begin{enumerate}
\item Unpack and install both the succeeding and failing versions of Coq and corresponding developments.
\item Replace the Coq binaries with wrappers that print out the arguments that Coq was called with, as well as \texttt{COQPATH} (an environment variable listing directories to be searched for imported modules) and the current directory.
\item Run Coq on the succeeding and failing developments, ensuring that the version that should pass does in fact pass, and the version that should fail has a recognizable error message.
\item Parse the build log to determine the buggy file name and the arguments to pass to Coq, using the extra logging introduced by our wrappers.
  This workflow means that we need not interface directly with varied build systems of different contributions on the CI.
\item Run Coq on the buggy file.
\item Parse the error message, ensuring that it matches with the error message from the build log.
  (See \autoref{sec:error-messages} for subtleties in that comparison)
\end{enumerate}

Again, the goal of the minimizer is to take a CI development that succeeds on the tip of the master branch and fails on a given pull request (PR), emitting a small, standalone file that succeeds on master and fails in the same way on the PR.
In order to do so efficiently, we reuse the CI artifacts from Coq.
We download the prebuilt versions of Coq from master and from the tip of the PR.
From just these artifacts and the name of the failing CI development, we must assemble enough information to run the bug minimizer.
We replicate Coq's generic CI workflow to install Coq as well as any dependencies of this CI development, into different directories: one for the version of Coq expected to pass and another for the version of Coq expected to fail.
We also reuse Coq's generic CI workflow to figure out the error message and the failing file we want to minimize.

Let us justify the extra information that our Coq wrapper programs log.
We need \texttt{COQPATH} to ensure that we have the right search path for the dependencies of \texttt{coqc}, the command-line Coq compiler.
We need the command-line arguments so that we know what flags to tell the bug minimizer to pass to \texttt{coqc}.
Note that we \emph{must not} change relative paths to absolute ones when passing arguments along to \texttt{coqc}, because the output of \texttt{coqc} is sensitive to the difference between relative and absolute paths, so changes can muddle tests that are meant to produce output files (and did in the past, for example with \texttt{ci-elpi}).
We can locate the error message by looking for the last instance of \texttt{File "$f$", line $\ell$, characters $n$-$m$:} followed immediately by a line beginning with \texttt{Error}.
(Note that warning messages also emit the \texttt{File $\ldots$} line, but we do not want to catch warnings.)
We look for the last instance of the wrapper debug printout information that points at the same file, though, so long as we were careful always to build single-threadedly, we could instead just look for the most recent debug printout before the error message.

Given this information, we adjust the arguments so that we can tell the bug minimizer where the dependencies live both for the passing and failing versions of Coq.
We then pass this information to the bug minimizer:
\begin{itemize}
\item the location of the file to be minimized;
\item the log file containing the error message, which must match the error message that the minimizer believes the file produces;
\item the locations of the \verb|coqc|, \verb|coqtop|, and \verb|coq_makefile| programs for the tip of the PR;
\item the location of the \verb|coqc| program for the master branch;
\item the locations of the dependencies for both the passing and failing versions of Coq, parsed from the command-line arguments and from walking the directories in \texttt{COQPATH};
\item any arguments to \verb|coqc| that are neither naming dependency locations nor known to be both irrelevant to the processing of the file and counterproductive to the minimizer's operation (such arguments are \verb|-batch|, which applies only to \verb|coqtop|; \verb|-time|, which will only make logs of the minimizer much longer; and \verb|-noglob|, \verb|-dump-glob|, and \verb|-o|, which interfere with the generation of outputs used by the minimizer).
\end{itemize}

\section{An Alternative Usage Mode}\label{sec:alt-usage}

Up to this point, we have talked about using the Coq Bug Minimizer exclusively to minimize reverse-CI failures for debugging faulty changes in Coq.
Our tool can also be used to minimize test cases for newly found bugs in Coq.
In this mode, a bug reporter can write a shell script that invokes a \emph{single} version of Coq to produce buggy behavior on some Coq file, asking coqbot to produce a minimal example from this script.
When running in this mode, we place an additional constraint on the minimizer that the proof script generating the error message should be left untouched,
which allows bug reporters to write proof scripts such as
\begin{verbatim}
some_tactic; lazymatch goal with
| buggy_goal => fail 0 "bug remains"
| [ |- ?G ]  => fail 0 "bug disappeared!" G end.
\end{verbatim}
to customize the desired reproducing case, trusting that the entire file will not be minimized to something silly like \verb|Goal False. fail 0 "bug remains"|.

\section{Integration in Coq's CI and Evaluation of Results}\label{sec:evaluation}

\subsection{Triggering the Minimizer}

The Coq project uses a custom, multi-task bot to automate everyday tasks, including triggering CI and reporting its results to the GitHub repository~\cite{zimmermann:hal-03479327}. We have extended this bot to automatically propose and manage the minimization of failing test cases. The bot posts a comment to propose to run minimization when a PR has passed Coq's internal test suite but has failures with external projects, and these external projects have built successfully on the base commit (on the master branch).

If someone answers with a comment to trigger minimization, then the bot prepares a branch with all the information needed by the minimizer and pushes this branch to an external repository dedicated to running the minimizer. This triggers a GitHub Action workflow which will proceed with the minimization process. GitHub Action jobs have a 6-hour timeout, so by the limit, the bot answers back with the results of the minimization process. If the minimization was stopped because of the timeout, then the bot automatically restarts it by reusing the file obtained at the previous step.

\subsection{Research Questions}

To evaluate the usefulness of our bug minimizer, we investigate several research questions:
\begin{description}
\item[RQ1:] How often does the minimizer successfully produce a reduced test case from the CI failures it was triggered on?
\item[RQ2:] How often is this reduced test case fully standalone (no dependencies other than Coq's standard library)?
\item[RQ3:] How long does it take to produce such reduced test cases?
\item[RQ4:] What is the size of the reduced cases?
\item[RQ5:] How long do the reduced cases take to run?
\item[RQ6:] What is the amount of code reduction?
\end{description}

\subsection{Data Collection and Analysis}

To support reproducing the results, we provide our data collection and analysis code (as a Jupyter notebook) and our dataset (as a CSV file) in the supplementary materials.

We retrieve the runs of the bug minimizer by looking for PRs in the Coq GitHub repository with the words ``coqbot ci minimize'', and we fetch all comments from the bot (timestamp and body text) from these PRs using GitHub's GraphQL API.
We exclude PRs opened by the first author, as most of these PRs were for testing the minimizer integration and debugging issues.
When the minimizer is triggered, the bot answers with a comment ``I have initiated minimization \ldots'' or ``I am now running minimization \ldots'', providing the list of projects on which it is being run.
Then, when it finishes minimizing a project, it produces a comment with the minimized file.
This file starts with header comments containing useful information about the minimization process.
The comment may also contain ``interrupted by timeout, being automatically continued'' if the minimization process timed out and has to be restarted to go further, which the bot automatically does.
We ignore these comments, only looking for final reduction outputs.
Finally, the bot posts a comment starting with ``Error: Could not minimize file'' when it was not able to minimize the requested failure, for instance, because it could not reproduce it or could not reproduce the successful run on the base branch.

We match comments indicating the start of the minimization with comments indicating the end of it, using these two comments to determine if the minimizer was able to produce a reduced test case, find how long it took, and answer our other research questions.
To avoid double-counting multiple runs on the same CI failure, we only look at the first bug-minimizer trigger on a given PR and a given project.

\subsection{Results}

\subsubsection{RQ1: How often does the minimizer produce a reduced test case?}

Looking only at the first minimization runs for a given PR and project, we have identified 191 runs on 51 PRs (very often, several minimization runs are started in the same PR on different projects).
On these 191 runs, 75\% succeeded in producing reduced test cases. We count as failed runs the ones where the bot reported ``Error: Could not minimize file'', the ones where we could not find a comment marking the end of minimization, and the ones where the bot answered with a minimized file but this file was not actually reduced from the initial test case (which we can detect from the header comments).

There were 5 runs for which we found no comment marking the end of minimization. By manually looking at them, we have determined that 4 out of 5 were caught in infinite loops and had to be canceled manually. Loops can arise when the 6-hour timeout of the minimization process is not enough to make any new progress and thus the minimization gets stuck without ever reaching its end. Typical circumstances are when testing out a single change takes over 20 seconds, since we only have enough time to compile a 20-second-long file about a thousand times in six hours. The last case of our 5 seems to be coqbot having failed to post the comment marking the end of the minimization process.

There were 19 runs that concluded with an explicit ``Error: could not minimize file'' comment. These errors are often due to issues downloading CI artifacts (9 runs), for instance because the corresponding base CI jobs have been skipped or the CI artifacts have expired. Runs concluding with errors can also happen because of bugs in Coq or in the tested projects' build infrastructure that prevent minimization. Virtually all these issues were reported, and most of them are already fixed. For instance, the MetaCoq project alone was responsible for 5 failures because of issues in its build system.

Finally, there were 23 runs ending with comments reporting on supposedly minimized files but where (from the header comments or their absence thereof) we can conclude that the minimization process failed to start properly (e.g., because it could not reproduce the error message). Most of these problems were related to error-message parsing, namespace management, or similar issues that have been fixed by making the bug minimizer more robust to them (see \autoref{sec:error-messages} to \autoref{sec:standalone}). A few of these issues have been noted but not yet fixed. Finally, a few of these failed runs were due to the minimizer being misused or called on a project that had failed for a reason that was unrelated to the PR.

The accompanying notebook contains specific comments for each of the failed runs.

\subsubsection{RQ2: How often is this reduced test case fully standalone?}

We consider that a reduced test case will be most useful if any dependency beyond Coq's standard library was successfully inlined, leaving it possible to run the reduced test case without needing to import any additional dependency. As a result, it is more likely that the test case can be added to Coq's test suite.

To measure how often the reduced test case is standalone, we rely on the minimizer recording when it failed to inline a dependency in the header comments of the minimized file. This feature was only added recently, so we only perform this measurement on the 47 successful runs of the minimizer that had this information available.
On these 47 runs, there were only 5 failures to inline dependencies fully, i.e., the minimizer produced a fully standalone file in 89\% of the cases.

Looking at the 5 failures to inline dependencies, we observe several types of reasons. One case was related to robustness to changing error messages, one case was related to a build-system issue in the project being minimized, and 3 cases were due to a common issue blocking attempts at all inlining methods. All of these issues have been fixed since then.

\subsubsection{RQ3: How long does it take to produce such reduced test cases?}

We compute the duration of minimization as the time delta between the start and the end comments.
This method overapproximates the actual time spent in the minimization process, since it also includes time setting up a VM and possibly waiting in the queue for an available runner.
We can look at this duration for both successful and failed runs.

For failed runs, we observe that the average duration for the minimization to conclude is 5 minutes (306 seconds) and that the maximum duration is 15 minutes (890 seconds).

For successful runs, we observe more variety. The minimum duration is 4 minutes (232 seconds), the maximum duration is 20 hours (73072 seconds), and the average is 104 minutes (6238 seconds). 50\% of the successful runs finish in under 20 minutes (1218 seconds), and 80\% finish in under 140 minutes (8396 seconds). This number is still reasonable compared to the time that contributors routinely spend waiting for the results of Coq's CI~\cite{zimmermann:tel-02451322}.

\subsubsection{RQ4: What is the size of the reduced cases?}

For the last three questions, we focus mainly on the 42 recent minimization runs that are known to have produced standalone files.

The shorter the reduced test case, the more useful it is: it can help developers understand the problem more quickly, and it makes it more likely that it will be added to Coq's test suite. Here again, there is some variety in the size of the reduced cases (counted in number of lines). The average size is 270 lines, and the maximum size is 2648 lines. However, 25\% of the reduced cases are under 39 lines, 50\% are under 114 lines, and 75\% are under 262 lines.

Results on the full set of 144 successful mininimization runs are of the same order of magnitude, with an average at 367 lines and a maximum size of 3804 lines.

Developers have the option to perform additional minimization manually and restart the automatic minimization process on their manually reduced cases, which can help obtain even more reduced cases, but we have not evaluated this feature quantitatively.

\subsubsection{RQ5: How long do the reduced cases take to run?}

Following a recent addition, the minimizer has reported the expected \texttt{coqc} compile time as part of the header comments in the minimized file. Our recent 42 standalone cases all had this field available. We observe that the reduced cases take on average 1.25 seconds to run, although 75\% of them take under half-a-second, while the maximum time is 26.5 seconds.

\subsubsection{RQ6: What is the amount of code reduction?}

To compute how much code reduction there was, we use data that the minimizer records about each minimization step (how many lines it started from and how many lines it ended up with). These numbers go up at times because of the process of inlining external dependencies. On the other hand, dependencies are only inlined if they could not simply be removed, so these numbers do not include the size of the files that were previously imported but did not need to be inlined during the minimization process.

We aggregate these numbers by simply taking the sum of the differences in line count at the beginning and the end of each minimization step. We compute the amount of code reduction by taking the ratio of the final size over the total test-case size, defined as being the sum of the final size and the total number of removed lines. We obtain an average figure of 31\%, which means that the final test-case size is on average one-third the size of the original test (including the dependencies that actually matter for the test case).

If we compute the size difference only looking at the initial file we started from and the final file we obtained, without accounting for the inlined dependencies, then we get an average ratio of 50\%, which means that the final file is on average half the size of the file we started from. Note that because of dependency inlining, nothing prevents the reduced test case from being longer than the file we started from, which does happen in 6 out of 42 cases. If we look only at the 36 cases for which there was some code reduction, we get that the average reduction is by a factor of 4 to 5. If we look only at the 6 cases for which there was code expansion, we get that the average expansion is by a factor of 2.

\subsection{Limitations of our Evaluation}

Evaluating a bug minimizer for a proof assistant such as Coq is difficult because there is no preexisting benchmark that it could be run on. In this paper, we have decided to take advantage of the integration of our minimizer in the CI infrastructure of Coq to evaluate it on real use cases where Coq developers have felt the need for it.

While we have taken steps to ensure that the evaluation is as unbiased as possible (such as not using reruns of the  minimization on the same project in the same PR), our evaluation is still limited by our choice to use real use cases. In particular, it should be noted that our evaluation results are not obtained on a fixed version of the minimizer. On the contrary, the minimizer has evolved (and is still evolving) in reaction to the very same cases on which we have evaluated it.
Since we always account only for first runs, many cases where the minimizer has been counted as failing have been eventually fixed and would result in successful runs today. Subsequent runs on other projects or other PRs may have succeeded thanks to earlier fixes.

Other limitations are that our computation of the minimization duration is an overapproximation that also includes the time for things such as setting up a VM to run the process, and that our evaluation of several research questions is based only on a subset of recent minimizer runs.

Due to all these limitations, our evaluation should only be understood as demonstrating the feasibility of our approach and the usefulness of its application to the development of Coq. However, it should not be understood as a basis that future versions of the minimizer, or alternative minimizers, can compare to, since today's version would already obtain different results if it were rerun on all these cases.

\section{Related Work}\label{sec:related-work}

Our work is at the crossing of two research areas: research on debugging techniques, which is a subdomain of software-engineering research, and research on proof assistants.

Debugging is a largely explored topic, but mostly with a focus on more mainstream and less formal languages than Coq.
In this research domain, test-case-reduction techniques have been studied for standard programming languages and compilers~\cite{chen_survey_compiler_testing}. There are two types of approaches that have been proposed. First, there are generic approaches that are supposed to work for any programming language, by using structure information on the program being reduced. Examples include delta debugging~\cite{Zeller2002} but also the generalized tree-reduction algorithm~\cite{herfert2017automatically} and the syntax-guided Perses tool~\cite{herfert2017automatically}. These generic techniques would not be likely to work well for Coq programs without careful adaptation, because many Coq programs can be considered syntactically valid even if completely nonsensical. For instance, we have already mentioned the issue with removing a \texttt{Qed} statement at the end of a tactic-based proof. Despite breaking a semantic block of code, this change does not actually produce a syntactically invalid Coq program.

Second, there are programming-language-specific approaches, which take advantage of specific knowledge to make the test reduction more performant. Our own work is related to this second category, where most tools focus on mainstream languages like C. Some are even dedicated to reducing the output of specific test-generation frameworks such as Csmith~\cite{regehr2012test}.

However, work on generating many diverse test cases from nothing has complementary value.
Csmith~\cite{yang2011finding} has an effective algorithm based on knowledge of C semantics, to provoke undefined behavior.
Techniques like equivalence modulo inputs~\cite{emi} find compiler bugs via differential testing, where a compiler is run on programs that are known to have the same semantics.
Perhaps this generative approach would also be useful for proof assistants, composed fruitfully with test-case reduction as we have presented.

Finally, the literature has identified the issue of \emph{slippage} in test reduction~\cite{chen2013taming,holmes2016mitigating}, which is when the initial and reduced cases produce different compiler bugs. This challenge was one of the main ones we had to account for in designing our bug minimizer (see \autoref{sec:error-messages}).

Proof-assistant ecosystems were already no stranger to testing techniques.
For instance, Isabelle/HOL's Nitpick~\cite{Nitpick} uses Boolean satisfiability to find theorem counterexamples.
QuickChick~\cite{quickchick} does random test generation to try to falsify Coq theorems.
These tools are handy to save users from investing time in trying to prove false theorems.
Testing-based approaches to debugging \emph{proof assistants themselves} are a complementary topic.

\section{Future Work}\label{sec:future-work}

We were pleasantly surprised to find that several ``shortcuts'' in the logistics behind the minimizer led to good results empirically, but some of these may be worth revisiting to improve results even more.
In various places, we use workarounds (like \texttt{.glob} files) to avoid integrating a proper Coq parser, but there would be advantages like being able to remove specific fields from record types.
We remove single commands at a time, rather than removing entire well-balanced command blocks, which probably costs us in minimization time.

A broader opportunity is finding related groups of commands that need to be removed together, to avoid changing the error message.
For instance, we might want to move a lemma out of a module, to the top level of a file.
Removing the commands that open and close the module might suffice, even if removing either one alone disturbs the error message.
A general-enough version of this process could replace many specific passes.

One remaining aggravation is proper handling of lemma proofs within sections, where the details of the lemma proof influence which section variables are kept in the lemma's type.
We could use the \verb|Set Suggest Proof Using| command to insert \verb|Proof using| clauses.

As mentioned in \autoref{sec:linearize-global-state}, we would like to improve the ability of the minimizer to linearize dependency trees and handle Coq's global state.
We could, for example, print out the full table of flag settings at a particular point, reset them to the initial values before inlining a file, and then restore them after inlining.
To fully handle global state, we would need some way to reconstruct the command-line flags used to compile installed files.

There are further-out ideas that could speed minimization significantly but might require significant modifications to Coq itself.
Incremental compilation would be helpful, to save us from rerunning long proof scripts every time we change single lines below them.
Minimizing multiple files in parallel, rather than only inlining files, would allow us to take advantage of multicore execution within single minimization jobs.

\nocite{coqpl-15-coq-bug-minimizer}
\bibliography{coq-bug-minimizer.bib}

\end{document}